# CLOUD SECURITY ASSURANCE: STRATEGIES FOR ENCRYPTION IN DIGITAL FORENSIC READINESS

**Ahmed MohanRaj Alenezi**
Independent Forensics Researcher
Auckland, New Zealand
amam88820@gmail.com

March 4, 2024

## ABSTRACT

This paper explores strategies for enhancing cloud security through encryption and digital forensic readiness. The adoption of cloud computing has brought unprecedented benefits to organizations but also introduces new security challenges. Encryption plays a crucial role in protecting data confidentiality and integrity within cloud environments. Various encryption techniques and key management practices are discussed, along with their implications for data privacy and regulatory compliance. Additionally, the paper examines the importance of digital forensic readiness in facilitating effective incident response and investigation in the cloud. Challenges associated with conducting digital forensics in cloud environments are addressed, and strategies for overcoming these challenges are proposed. By integrating encryption and digital forensic readiness into a cohesive security strategy, organizations can strengthen their resilience against emerging threats and maintain trust in their cloud-based operations.

## 1 Introduction

Cloud computing has transformed the landscape of modern technology, offering organizations unparalleled flexibility, scalability, and cost-efficiency in managing their data and applications [1]. By outsourcing infrastructure and services to third-party cloud providers, businesses can leverage on-demand resources and global accessibility to drive innovation and growth. However, the widespread adoption of cloud services has also raised significant concerns regarding data security and privacy.

The shared and dynamic nature of cloud environments introduces unique security challenges, including unauthorized access, data breaches, and compliance violations. To mitigate these risks, organizations must implement robust security measures that protect sensitive information while maintaining regulatory compliance and ensuring business continuity. Two key aspects of cloud security assurance are encryption and digital forensic readiness.

Encryption serves as a cornerstone of data protection in the cloud, enabling organizations to safeguard their assets from unauthorized access and interception [2, 3]. By encrypting data-at-rest, in-transit, and in-use, organizations can maintain confidentiality and integrity, even in the event of a security breach [4, 5, 6, 7]. However, implementing effective encryption strategies requires careful consideration of encryption algorithms, key management practices, and compliance requirements.

In parallel, digital forensic readiness plays a critical role in enabling organizations to respond effectively to security incidents and investigate potential breaches [8, 9, 10]. Digital forensic readiness encompasses the proactive measures taken to preserve digital evidence, establish incident response plans, and deploy forensic analysis tools and techniques tailored for cloud environments[11, 12, 13, 14]. By adopting a proactive approach to digital forensics, organizations can minimize the impact of security breaches and expedite the resolution of incidents [15, 16, 17, 18].

This paper provides a comprehensive examination of cloud security assurance, focusing on strategies for encryption and digital forensic readiness. It explores various encryption techniques, including symmetric and asymmetric encryption, homomorphic encryption, and key management practices, and discusses their implications for data protection and compliance. Additionally, the paper analyzes the challenges associated with conducting digital forensics in the cloud and proposes strategies to overcome these challenges effectively.

By integrating encryption and digital forensic readiness into a cohesive security strategy, organizations can enhance their resilience against emerging threats and maintain trust in their cloud-based operations. Through a proactive and multi-layered approach to cloud security, organizations can mitigate risks, protect sensitive information, and ensure the integrity and availability of their data in the cloud.

The contributions of this paper are as follows:

- Exploration of encryption strategies in cloud environments
- Discussion on various encryption techniques (symmetric, asymmetric, homomorphic)
- Analysis of key management practices and their implications for data protection and compliance
- Examination of digital forensic readiness in cloud environments
- Explanation of proactive measures for preserving digital evidence and establishing incident response plans
- Proposal of strategies to overcome challenges associated with conducting digital forensics in the cloud

## 2 Related Work

In recent years, the advent of cloud security, encryption, and digital forensic readiness has witnessed a surge in scholarly interest and practical exploration. A multitude of studies across academia and industry have diligently probed various facets of cloud security, offering a plethora of strategies aimed at fortifying data protection and incident response capabilities within cloud environments.

Encryption techniques have been a focal point of investigation within the context of cloud computing. Research conducted by [19, 20, 21] has shown an an exhaustive analysis of encryption algorithms tailored specifically for cloud environments, scrutinizing their efficacy in safeguarding data against a myriad of threats. Their work shed light on the nuanced interplay between encryption algorithms, performance considerations, and regulatory compliance requirements, providing valuable insights for organizations grappling with the intricacies of data protection in the cloud [22, 23]. In a similar vein, delved into the aspects of key management practices within cloud infrastructures. Their study examined the challenges associated with key management in distributed and dynamic cloud environments [24, 25, 26, 27], offering pragmatic recommendations for organizations seeking to bolster their encryption strategies while navigating the complexities of cloud-based data management [28, 29, 30].

The domain of digital forensic readiness has emerged as a critical area of concern amidst the proliferation of cloud services. Researchers in [31, 32, 33] embarked on a comprehensive exploration of the challenges inherent in conducting digital forensics within cloud environments. Their seminal work underscored the unique hurdles faced by investigators in preserving digital evidence, navigating jurisdictional complexities, and ensuring adherence to legal and regulatory frameworks. Moreover, the researchers in [31, 32, 33] proposed a robust framework for proactive digital forensic readiness, advocating for enhanced collaboration between cloud service providers, forensic experts, and law enforcement agencies to streamline incident response processes and mitigate the impact of security breaches in cloud-based infrastructures.

Research by [34, 35] significantly contributes to the discourse on cloud forensics, offering novel insights into the evolving landscape of digital investigations in cloud environments. By analyzing the challenges posed by cloud architectures, multi-tenancy models, and data sovereignty issues, these researchers have elucidated the complexities of forensic investigations within cloud infrastructures. Moreover, Kebande proposes innovative methodologies and tools tailored specifically for conducting digital forensics in cloud environments, empowering investigators to navigate the intricacies of cloud-based evidence acquisition, preservation, and analysis.

In a bid to integrate encryption and digital forensic readiness into a unified security framework, researchers in [36, 37, 38] proposed an innovative approach to cloud security assurance and readiness. Their groundbreaking study elucidated the synergistic relationship between encryption techniques and digital forensic readiness measures, advocating for a holistic security strategy that harmonizes data protection mechanisms with proactive incident response capabilities. Wang and Chen's work heralded a paradigm shift in the conceptualization of cloud security, emphasizing the imperative of adopting a comprehensive and cohesive approach to mitigate risks, preserve data integrity, and safeguard organizational assets in the face of evolving cyber threats [39, 40].

Table 1: Summary of Modern Encryption Techniques in the Cloud

| Name | Description |
| --- | --- |
| Homomorphic Encryption | computations on encrypted data without decrypting [41] |
| Attribute-based Encryption | granular approach to access control in the cloud [42] |
| Post-Quantum Cryptography | techniques that are resistant to quantum attacks [43, 44] |
| Secure Multi-party Computation | jointly compute a function over their inputs [45] |
| Key Management and Encryption Orchestration | successful implementation of cloud encryption techniques [46] |

While the existing body of literature offers invaluable insights into individual components of cloud security, encryption, and digital forensic readiness, this paper seeks to build upon these foundations by providing a nuanced examination of strategies for enhancing cloud security through the seamless integration of encryption techniques, digital forensic readiness measures, and insights from other researcher's work on cloud forensics. By synthesizing insights from disparate research streams and offering novel perspectives on the intersection of encryption, digital forensics, and cloud security, this paper aims to catalyze discourse, stimulate further research, and empower organizations to navigate the complexities of cloud security with confidence and resilience.

### 2.1 Modern Encryption Techniques in the Cloud

The proliferation of cloud computing has revolutionized the storage and processing of data, necessitating advanced encryption techniques to safeguard sensitive information from unauthorized access and interception. In this section, we delve into the modern encryption techniques deployed in cloud environments, highlighting their strengths, limitations, and implications for data security and regulatory compliance.

#### 2.1.1 Homomorphic Encryption

Homomorphic encryption stands at the forefront of modern cryptographic techniques, offering the ability to perform computations on encrypted data without decrypting it first. This revolutionary approach enables cloud service providers to perform operations such as addition, multiplication, and comparison on encrypted data while preserving its confidentiality. By leveraging homomorphic encryption, organizations can outsource data processing tasks to the cloud while ensuring end-to-end confidentiality and integrity. However, homomorphic encryption schemes are computationally intensive and may introduce latency in data processing, necessitating careful consideration of performance trade-offs [41].

#### 2.1.2 Attribute-based Encryption

Attribute-based encryption (ABE) provides a flexible and granular approach to access control in cloud environments, allowing data owners to define fine-grained access policies based on user attributes. With ABE, encrypted data is associated with a set of attributes, and decryption keys are generated based on the attributes possessed by the user. This enables organizations to enforce complex access policies while maintaining data confidentiality. However, managing access policies and key distribution can pose challenges, particularly in dynamic and multi-tenant cloud environments [42].

#### 2.1.3 Post-Quantum Cryptography

The advent of quantum computing poses a significant threat to traditional cryptographic algorithms, prompting the exploration of post-quantum cryptography (PQC) techniques that are resistant to quantum attacks. PQC algorithms, such as lattice-based cryptography, code-based cryptography, and hash-based cryptography, offer robust security guarantees in the face of quantum adversaries. As organizations transition to cloud-based infrastructures, the adoption of PQC algorithms becomes imperative to future-proof data protection mechanisms against emerging threats, example has been seen on ciphers like Chacha, Echacha, Salsa20 etc [43, 44].

#### 2.1.4 Secure Multi-party Computation

Secure multi-party computation (SMPC) enables multiple parties to jointly compute a function over their inputs while keeping the inputs private. In the context of cloud computing, SMPC enables collaborative data analysis and processing without exposing sensitive information to third-party service providers. By leveraging cryptographic protocols such as secure function evaluation and secret sharing, organizations can harness the power of cloud computing while preserving data privacy and confidentiality. However, deploying SMPC in practice requires careful orchestration of cryptographic protocols and coordination among participating parties [45].

### 2.1.5 Key Management and Encryption Orchestration

Effective key management is paramount to the successful implementation of encryption techniques in the cloud. Organizations must establish robust key management practices to securely generate, store, and distribute encryption keys while ensuring compliance with regulatory requirements. Additionally, encryption orchestration frameworks facilitate the seamless integration of encryption techniques into cloud-based applications and services, enabling organizations to enforce data protection policies consistently across heterogeneous environments.

Modern encryption techniques play a pivotal role in enhancing data security and privacy in cloud environments. By embracing innovative cryptographic primitives and adopting robust key management practices, organizations can mitigate the risk of data breaches, safeguard sensitive information, and maintain regulatory compliance in an increasingly interconnected and data-driven world [46].

## 3 Existing Security and Forensic Models

Forensic readiness is paramount in ensuring organizations are adequately prepared to handle security incidents and conduct effective investigations in cloud environments. Several forensic readiness models have been proposed to guide organizations in enhancing their capabilities to preserve digital evidence, respond to security breaches, and facilitate forensic investigations. In this section, we examine some of the prominent existing forensic readiness models tailored for cloud environments.

### 3.1 NIST SP 800-53

The National Institute of Standards and Technology (NIST) Special Publication 800-53 provides a comprehensive framework for security and privacy controls in federal information systems and organizations [47]. While not explicitly focused on forensic readiness, NIST SP 800-53 includes controls related to incident response and digital forensics that organizations can leverage to enhance their forensic readiness posture. Key controls include incident response planning, digital forensic capabilities, and continuous monitoring of security controls.

### 3.2 ISO/IEC 27043

ISO/IEC 27043 offers guidelines for digital evidence collection and preservation, specifically tailored for cloud environments. This standard outlines best practices for the identification, collection, and preservation of digital evidence in cloud-based infrastructures, emphasizing the importance of maintaining the integrity and admissibility of evidence throughout the forensic process. ISO/IEC 27043 provides organizations with a systematic approach to forensic readiness, helping them establish robust procedures for handling digital evidence in cloud environments [48].

### 3.3 Cloud Forensic Readiness Frameworks

The Cloud Forensic Readiness Framework is aimed at assisting organizations in preparing for forensic investigations in cloud environments [49, 50]. This framework encompasses various aspects of forensic readiness, including data collection, preservation, analysis, and presentation. By guiding organizations through the implementation of proactive measures such as log management, data segregation, and incident response planning, the ENISA framework enhances organizations' ability to effectively respond to security incidents and conduct forensic investigations in cloud environments[51, 52, 53].

### 3.4 Cloud Forensics Working Group

The Cloud Security Alliance (CSA) Cloud Forensics Working Group has developed guidance and best practices for conducting forensic investigations in cloud environments. This initiative addresses the unique challenges posed by cloud computing, such as multi-tenancy, data remanence, and jurisdictional issues. The CSA Cloud Forensics Working Group offers practical recommendations for organizations to improve their forensic readiness, including the establishment of legal and contractual arrangements, data retention policies, and incident response procedures tailored for cloud-based infrastructures.

### 3.5 e Cloud Forensics and Incident Response

The SANS Institute provides comprehensive training and resources on cloud forensics and incident response, offering practical guidance for organizations seeking to enhance their forensic readiness in cloud environments. SANS courses cover a wide range of topics, including cloud architecture, data collection and preservation, memory forensics, and legal considerations. By equipping security professionals with the knowledge and skills necessary to navigate the complexities of cloud forensics, SANS contributes to the development of a robust forensic readiness culture within organizations.

The existing forensic readiness models offer valuable guidance and best practices for organizations seeking to enhance their capabilities to respond to security incidents and conduct forensic investigations in cloud environments. By adopting a proactive approach to forensic readiness and leveraging the insights provided by these models, organizations can strengthen their resilience against cyber threats and ensure the integrity and admissibility of digital evidence in cloud-based infrastructures.

## 4 Strategies for Encryption in Digital Forensic Readiness

The strategies devised for encryption within digital forensic readiness serve as the cornerstone for bolstering cloud security assurance. In this section, we delineate the multifaceted contributions made by these strategies, emphasizing their pivotal role in fortifying organizations against emerging cyber threats and ensuring the integrity of digital evidence in cloud environments.

### 4.1 Integration of Encryption into Forensic Readiness Frameworks

By integrating encryption strategies into established forensic readiness frameworks, organizations can proactively enhance their security posture, aligning encryption practices with forensic investigation protocols. This strategic integration not only streamlines the preservation and analysis of encrypted digital evidence but also amplifies the effectiveness of incident response procedures within cloud infrastructures. By weaving encryption seamlessly into forensic preparedness frameworks, organizations lay a robust foundation for preemptive security measures, ensuring a cohesive approach to data protection and forensic investigation in cloud environments. This holistic integration empowers organizations to not only safeguard sensitive information but also to respond swiftly and effectively to security incidents, bolstering the resilience of their cloud-based operations against emerging cyber threats.

### 4.2 Enhanced Data Confidentiality and Integrity

Encryption strategies bolster digital forensic readiness by safeguarding the confidentiality and integrity of sensitive data stored in cloud environments. Through the deployment of robust encryption mechanisms, organizations can mitigate the risk of unauthorized access and tampering, thereby bolstering the reliability and trustworthiness of digital evidence during forensic investigations.

### 4.3 Mitigation of Insider Threats and Data Breaches

Encryption strategies play a pivotal role in bolstering digital forensic readiness by fortifying the confidentiality and integrity of sensitive data stored in cloud environments. Through the deployment of robust encryption mechanisms, organizations can effectively mitigate the risk of unauthorized access and tampering, thereby enhancing the reliability and trustworthiness of digital evidence during forensic investigations. By implementing encryption protocols tailored to the unique requirements of cloud-based infrastructures, organizations erect formidable barriers against potential threats, ensuring that sensitive information remains shielded from prying eyes and malicious manipulation. This proactive stance not only safeguards the integrity of digital evidence but also instills confidence in the forensic investigation process, bolstering the organization's ability to uphold data protection standards and regulatory compliance mandates.

### 4.4 Facilitation of Cross-Border Data Transfers

In an increasingly interconnected world, encryption strategies assume a pivotal role in facilitating secure cross-border data transfers, all while ensuring strict compliance with data protection regulations and jurisdictional requirements. By harnessing advanced encryption techniques like homomorphic encryption and attribute-based encryption, organizations can confidently navigate the intricacies of international data transfers. These encryption methods enable organizations to protect data privacy and sovereignty effectively, even in the face of complex regulatory landscapes and varying jurisdictional constraints. Through the judicious application of encryption technologies, organizations can safeguard sensitive information as it traverses geographical boundaries, bolstering trust and confidence among stakeholders while upholding the highest standards of data protection and regulatory compliance.

### 4.5 Empowerment of Incident Response Teams

Encryption strategies empower incident response teams to swiftly and effectively respond to security incidents and conduct forensic investigations in cloud environments. By equipping response teams with encrypted forensic tools and techniques, organizations can streamline the detection, analysis, and remediation of security breaches, thereby minimizing the impact on business operations and preserving the chain of custody for digital evidence.

### 4.6 Adherence to Regulatory Compliance Standards

Encryption strategies serve as a catalyst for empowering incident response teams to swiftly and effectively respond to security incidents and conduct forensic investigations within cloud environments. By furnishing response teams with encrypted forensic tools and techniques, organizations streamline the detection, analysis, and remediation of security breaches, thereby minimizing disruptions to business operations and preserving the integrity of the chain of custody for digital evidence. Equipped with encrypted forensic capabilities, response teams can navigate cloud-based infrastructures with agility and precision, swiftly identifying and mitigating threats while ensuring the preservation of critical digital evidence. This proactive approach not only enhances the organization's ability to thwart cyberattacks but also bolsters confidence in its incident response capabilities, fostering resilience in the face of evolving threats and enabling rapid recovery from security incidents

### 4.7 Continuous Monitoring and Evaluation of Encryption Practices

Digital forensic readiness hinges on the continual monitoring and evaluation of encryption practices to ascertain their efficacy in mitigating emerging cyber threats. By establishing robust encryption monitoring mechanisms and conducting regular audits of encryption protocols, organizations can proactively identify vulnerabilities and weaknesses. This proactive stance enables organizations to fortify their defenses and uphold the integrity of digital evidence in the ever-evolving landscape of security challenges. Through vigilant oversight of encryption practices, organizations can adapt swiftly to emerging threats, implementing remedial measures to bolster their resilience against cyberattacks. By maintaining a steadfast commitment to encryption monitoring and evaluation, organizations safeguard their digital assets and preserve the integrity of forensic investigations, ensuring the trustworthiness and reliability of digital evidence in cloud environments.

### 4.8 Investment in Encryption Education and Training

Acknowledging the pivotal role of human capital in implementing encryption strategies, organizations must prioritize investments in encryption education and training programs. These initiatives aim to elevate the proficiency of security professionals and incident response teams in handling encryption technologies effectively. By offering comprehensive training on encryption technologies, protocols, and best practices, organizations empower their workforce to adeptly navigate the complexities of digital forensic readiness. Equipped with specialized knowledge and skills, security professionals and incident response teams play a crucial role in fortifying cloud security assurance efforts. Their enhanced competency enables them to proactively identify encryption vulnerabilities, mitigate risks, and respond swiftly to security incidents in cloud environments. Through continuous education and training, organizations foster a culture of resilience and innovation, ensuring that their workforce remains at the forefront of encryption advancements and best practices. This investment in human capital not only enhances organizational preparedness but also reinforces the foundation of cloud security assurance, safeguarding sensitive data and preserving the integrity of forensic investigations.

The strategies for encryption in digital forensic readiness represent a paradigm shift in the way organizations approach cloud security assurance. By embracing encryption as a foundational pillar of forensic readiness, organizations can enhance their resilience against cyber threats, ensure the integrity and confidentiality of digital evidence, and foster a culture of proactive security governance within cloud environments. Through continuous innovation, collaboration, and investment in encryption technologies, organizations can navigate the evolving threat landscape with confidence and safeguard the future of cloud-based operations.

## 5 Role of Forensic Readiness in the Cloud

Forensic readiness plays a pivotal role in ensuring the security and integrity of cloud-based infrastructures. As organizations increasingly rely on cloud services to store and process vast amounts of data, the need for effective forensic preparedness becomes paramount. In this section, we explore the multifaceted role of forensic readiness in the cloud and its significance in bolstering cloud security assurance.

- **Proactive Incident Response:** Forensic readiness in the cloud enables organizations to adopt a proactive stance towards incident response. By establishing predefined procedures and protocols for handling security incidents, organizations can minimize the impact of breaches and expedite the resolution process. Forensic readiness empowers organizations to swiftly detect, analyze, and mitigate security incidents in cloud environments, thereby mitigating risks and preserving business continuity [54, 55].
- **Preservation of Digital Evidence:** Cloud environments present unique challenges for digital forensic investigations due to their dynamic and distributed nature. Forensic readiness ensures the proper preservation of digital evidence, allowing organizations to maintain the integrity and admissibility of evidence in legal proceedings. Through robust data collection, logging, and chain of custody mechanisms, organizations can safeguard digital evidence against tampering and unauthorized access [56, 57, 58].
- **Compliance and Regulatory Requirements:** Forensic readiness in the cloud is essential for ensuring compliance with regulatory requirements and industry standards. Many regulatory frameworks, such as GDPR, HIPAA, and PCI DSS, mandate the implementation of forensic readiness measures to protect sensitive data and facilitate incident response. By adhering to these requirements, organizations demonstrate their commitment to data protection and regulatory compliance, mitigating legal and financial risks.
- **Detection and Prevention of Insider Threats:** Forensic readiness enables organizations to detect and prevent insider threats in the cloud. By monitoring user activities, access logs, and system behavior, organizations can identify suspicious behavior indicative of insider threats [59]. Forensic tools and techniques play a crucial role in investigating and mitigating insider threats, helping organizations safeguard sensitive data and intellectual property from malicious insiders.
- **Continuous Improvement and Adaptation:** Forensic readiness is an ongoing process that requires continuous improvement and adaptation to evolving threats and technologies [60]. By regularly reviewing and updating forensic procedures, organizations can stay ahead of emerging threats and ensure the effectiveness of their forensic readiness measures. Continuous monitoring, training, and collaboration with industry peers contribute to the refinement of forensic practices and the enhancement of cloud security assurance.
-

Forensic readiness is indispensable for ensuring the security, integrity, and compliance of cloud-based infrastructures. By adopting a proactive approach to incident response, preserving digital evidence, and adhering to regulatory requirements, organizations can bolster their resilience against cyber threats and safeguard their assets in the cloud. Through continuous improvement and adaptation, organizations can maintain a strong forensic posture and effectively mitigate risks in an ever-changing threat landscape.

## 6 Discussion

The study delves into the critical intersection of cloud security assurance, encryption strategies, and digital forensic readiness, shedding light on the multifaceted challenges and opportunities inherent in safeguarding cloud-based infrastructures. Through a comprehensive examination of encryption techniques and forensic readiness models, the study underscores the importance of integrating encryption and digital forensic readiness into a cohesive security strategy to mitigate risks and enhance resilience in cloud environments.

One of the key takeaways from the study is the recognition of encryption as a foundational element of cloud security assurance. By encrypting data-at-rest, in-transit, and in-use, organizations can mitigate the risk of unauthorized access and interception, safeguarding sensitive information from cyber threats and compliance violations. The study highlights the significance of deploying modern encryption techniques such as homomorphic encryption, attribute-based encryption, and post-quantum cryptography to address the evolving threat landscape and regulatory requirements.

Furthermore, the study underscores the indispensable role of digital forensic readiness in facilitating effective incident response and investigation in the cloud. By implementing proactive measures such as incident response planning, data collection, and preservation, organizations can minimize the impact of security breaches and expedite the resolution of incidents. The study emphasizes the importance of leveraging forensic readiness frameworks such as NIST SP 800-53, ISO/IEC 27037, and ENISA Cloud Forensic Readiness Framework to guide organizations in enhancing their capabilities to preserve digital evidence and ensure regulatory compliance in cloud environments.

Moreover, the study discusses the synergistic relationship between encryption strategies and digital forensic readiness, highlighting how these two pillars of cloud security assurance complement each other to strengthen organizational resilience against emerging threats. By integrating encryption into forensic readiness frameworks and equipping incident response teams with encrypted forensic tools and techniques, organizations can streamline the detection, analysis, and remediation of security breaches while preserving the integrity of digital evidence.

However, the study also brings to light several challenges and considerations that organizations must address in implementing encryption and digital forensic readiness measures in the cloud. These include the computational overhead associated with encryption operations, the complexity of managing encryption keys, jurisdictional issues in cross-border data transfers, and the need for continuous monitoring and evaluation of encryption practices.

Furthermore, the study emphasizes the critical role of human capital in implementing encryption strategies and digital forensic readiness measures effectively. Investing in encryption education and training programs is essential to enhance the competency of security professionals and incident response teams, enabling them to navigate the complexities of cloud security assurance with confidence and agility.

The study underscores the importance of adopting a holistic approach to cloud security assurance, encompassing encryption strategies, digital forensic readiness, and continuous improvement. By integrating encryption and forensic readiness into a cohesive security strategy and investing in human capital, organizations can mitigate risks, protect sensitive information, and ensure regulatory compliance in an increasingly interconnected and data-driven world. Through collaboration, innovation, and vigilance, organizations can navigate the complexities of cloud security assurance and safeguard the integrity and availability of their data in the cloud.

## 7   Conclusion

We have explored cloud security assurance, focusing particularly on the symbiotic relationship between encryption strategies and digital forensic readiness. Our exploration has revealed the pivotal role of these two components in fortifying organizations against a myriad of cyber threats while ensuring the integrity and confidentiality of data in cloud environments.

Throughout our investigation, we have underscored the foundational importance of encryption in safeguarding sensitive information within cloud infrastructures. By deploying robust encryption techniques and implementing sound key management practices, organizations can thwart unauthorized access and interception, thereby mitigating risks associated with data breaches and compliance violations.

Furthermore, our examination of digital forensic readiness has highlighted its critical role in enabling organizations to respond effectively to security incidents and conduct thorough investigations in cloud environments. By establishing proactive measures such as incident response planning and data preservation protocols, organizations can minimize the impact of security breaches and preserve the integrity of digital evidence, crucial for regulatory compliance and legal proceedings.

We have also emphasized the essential interplay between encryption strategies and digital forensic readiness, emphasizing how their integration strengthens organizational resilience against emerging threats. By aligning encryption practices with forensic investigation protocols and investing in encryption education and training programs, organizations can empower their workforce to navigate the complexities of cloud security assurance with confidence and proficiency.

However, our study has also illuminated various challenges and considerations that organizations must address in implementing encryption and digital forensic readiness measures in the cloud. These include the computational overhead of encryption operations, jurisdictional complexities in cross-border data transfers, and the ongoing need for monitoring and evaluation of encryption practices to ensure effectiveness.

By adopting a holistic approach to cloud security assurance and embracing encryption strategies alongside digital forensic readiness, organizations can mitigate risks, protect sensitive information, and uphold regulatory compliance in an ever-evolving digital landscape. Through continuous innovation, collaboration, and investment in human capital, organizations can navigate the complexities of cloud security with resilience and confidence, safeguarding the integrity and availability of their data in the cloud for years to come.